\documentclass[%
superscriptaddress,
twocolumn,
showpacs,preprintnumbers,
amsmath,amssymb,
aps,
prl,
]{revtex4-1}

\usepackage{graphicx}% Include figure files
\usepackage{dcolumn}% Align table columns on decimal point
\usepackage{bm}% bold math
\newcommand{\CFMO}{Ca$_{2}$FeMnO$_{6}$}
\newcommand{\CTMO}{Ca$_{2}$TiMnO$_{6}$}

\begin{document}

%\preprint{APS/123-QED}

\title{Unusual layered order and charge disproportionation
in double perovskite Ca$_{2}$FeMnO$_{6}$}
% Force line breaks with \\

\author{Ke Yang}
 \affiliation{Laboratory for Computational Physical Sciences (MOE),
 State Key Laboratory of Surface Physics, and Department of Physics,
 Fudan University, Shanghai 200433, China}%Lines break automatically or can be forced with \\

\author{D. I. Khomskii}
\affiliation{Institute of Physics II, University of Cologne, 50937 Cologne, Germany}

\author{Hua Wu}
\email{Corresponding author. wuh@fudan.edu.cn}
\affiliation{Laboratory for Computational Physical Sciences (MOE),
 State Key Laboratory of Surface Physics, and Department of Physics,
 Fudan University, Shanghai 200433, China}
\affiliation{Collaborative Innovation Center of Advanced Microstructures,
 Nanjing 210093, China}

%\affiliation{
%\affiliation{
%\affiliation{%

\date{\today}

\begin{abstract}
While double perovskites A$_{2}$BB'O$_{6}$, if ordered, usually form a rock-salt-type structure with a checkerboard B/B' ordering, it is surprising that {\CFMO} has alternate FeO$_{2}$ and MnO$_{2}$ layers in its perovskite structure. Here we demonstrate, using density functional calculations, that this unusual layered ordering facilitates, and is largely helped by, the Fe$^{3+}$-Fe$^{5+}$ charge disproportionation (CD) of the formal Fe$^{4+}$ ions, which would
otherwise be frustrated in the common rock salt structure. To further verify the important role of the CD for stabilization of this layered ordering, we carry out a comparative study for the isostructural Ca$_{2}$TiMnO$_{6}$ which has a simple Ti$^{4+}$-Mn$^{4+}$ state free of the CD. Our calculations indicate that Ca$_{2}$TiMnO$_{6}$ instead prefers the standard rock salt structure to the layered one. Thus our study shows a nontrivial interplay between the CD and the type of ion ordering, and proves that the CD is strongly involved in stabilizing the unusual layered order of {\CFMO}.

%\begin{description}
%\end{description}
\end{abstract}

\maketitle

%\tableofcontents

\section{\label{sec:lint}Introduction}
%%introduction

When dealing with transition metal compounds one often looks at different degrees of freedom like charge, spin and orbital and their interplay with the lattice.~\cite{Tokura,Khomskii}
An interesting phenomenon--charge disproportionation (CD)~\cite{Alonso,Mazin,Kawasaki,Woodward,Yamada,Mizumaki} occurs in the materials with a nominal integer valence which is however intrinsically unstable. In such systems there occurs a spontaneous charge segregation at low temperatures, for example, CaFeO$_{3}$ shows a CD (2Fe$^{4+}$ $\rightarrow$ Fe$^{3+}$ + Fe$^{5+}$) at 290 K.~\cite{Kawasaki,Woodward} A similar CD transition at 210 K is observed in the A-site (Ca,Cu) ordered perovskite CaCu$_{3}$Fe$_{4}$O$_{12}$ with the nominal Fe$^{4+}$.~\cite{Yamada,Mizumaki} Below the CD transition temperature, the charge-disproportionated Fe$^{3+}$ and Fe$^{5+}$ ions at the B sites in a perovskite structure are usually ordered in a rock-salt manner.~\cite{Kawasaki,Woodward,Yamada,Mizumaki} (Note that for the high valent transition metal ions and particularly those with a negative charge transfer energy,\cite{Khomskii,Zaanen,Bocquet} actually a large fraction of charges are here on ligand oxygens, i.e. for example Fe$^{5+}$ is rather Fe$^{3+}$$\underline{L}^2$, where $\underline{L}$ is a ligand hole, see e.g. Refs. ~\cite{Khomskii,Zaanen,Bocquet} This however is not crucial for the further discussion, thus we will continue to speak about Fe$^{5+}$ -- of course keeping in mind this remark).

The CD is observed structurally via the cooperative lattice distortion which accommodates the CD ions in different charge state and thus with different ionic size and bond length.\cite{Alonso,Mazin,Woodward,Mizumaki} It can also be  identified by the M\"ossbauer spectrum.\cite{Hosaka17,Hosaka15} Moreover, this transition is often simultaneously accompanied by a metal-insulator transition: the formation of Fe$^{3+}$-Fe$^{5+}$ superstructures leads to a gap opening in the electronic spectrum.\cite{Kawasaki,Yamada} A similar situation is observed in rare earth nickelate perovskites RNiO$_{3}$. The phase diagram of this family of compounds shows that in most of them there occurs a metal-insulator transition with decreasing temperature, accompanied or driven by the formation of CD, formally 2Ni$^{3+}$ $\rightarrow$ Ni$^{2+}$ + Ni$^{4+}$.~\cite{Alonso,Mazin,Mizokawa,Johnston}

Very recently, the double perovskite {\CFMO} was prepared and it has the nominal Fe$^{4+}$ ($t_{2g}^3e_g^1$) and Mn$^{4+}$ ($t_{2g}^3$) ions.\cite{Hosaka15} Again, the Fe$^{4+}$ ions are unstable against CD. In the ordered phase of this material, surprisingly the Fe and Mn ions are ordered in alternate layers but not in the common rock salt structure, see Fig. 1. Moreover, this material undergoes a CD transition below 200 K, forming a checkerboard arrangement of the formal Fe$^{3+}$ and Fe$^{5+}$ ions in each FeO$_2$ layer. As mixed B-site ions may be disordered in perovskites when their ionic sizes are not much different,\cite{Anderson,Serrate} it is interesting that here the B-site Fe$^{4+}$ (0.585~\AA) and Mn$^{4+}$ (0.530~\AA)\cite{Shannon} ions in a similar size are ordered at all, not even to speak about the surprising layered ordering, uncommon in mixed double perovskites. This is the main motivation for us to carry out detailed first principles calculations to address this issue.

As seen below, our results show that this unusual layered ordering facilitates the occurrence of the Fe$^{3+}$-Fe$^{5+}$ CD of the formal Fe$^{4+}$ ions, which however would be frustrated in the standard rock salt structure. In its turn, the Fe$^{4+}$ CD tendency actually stabilizes this surprising layered ordering, i.e. these two phenomena are intrinsically interrelated.
To substantiate our conclusion, we also carry out a comparative study on the isostructural Ca$_{2}$TiMnO$_{6}$ with the CD free Ti$^{4+}$ and Mn$^{4+}$ ions. Our calculations show that Ca$_{2}$TiMnO$_{6}$  would prefer the common rock salt structure to the layered one. All this demonstrates that the CD is indeed involved in forming the unusual layered ordering of Ca$_{2}$FeMnO$_{6}$.

 \begin{figure}[t]
\includegraphics[width=7cm]{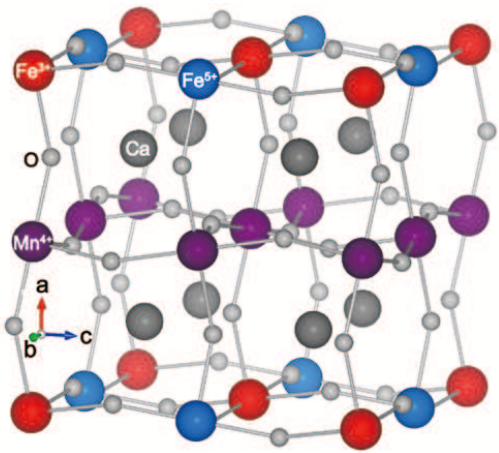}
 \caption{Crystal structure of the double perovskite {\CFMO} with the unusual Fe/Mn layered order.
}
%\label{fig2:el}
\end{figure}
%

%%method
\section{\label{sec:method}Computational Details}

The density functional calculations were performed using the full-potential augmented plane wave plus local orbital code (Wien2k).~\cite{WIEN2K} The $2\times2\times2$ supercell of the ABO$_3$ perovskite type was set for both the layered structure (see Fig. 1) and rock salt one. The experimental lattice parameters\cite{Hosaka15} were used and the structural optimization was also carried out. The muffin-tin sphere radii was chosen to be 2.5, 2.0, 2.0 and 1.3 Bohr for Ca, Fe, Mn, and O, respectively. The plane wave expansion of the interstitial wave functionals was set to be 15 Ry.  The Brillion zone integration in the course of self-consistent iterations was performed over 5$\times$5$\times$5 mesh in K-space. The typical value of Hubbard U=5.0 eV and Hund exchange J=1.0 eV were used for both Fe and Mn 3d states in the LSDA+U calculations to account for the electron correlations.~\cite{Anisimov} A same set of computational parameters was used in the comparative study of the isostructural Ca$_2$TiMnO$_6$.

\section{Results and Discussion}

We started with LSDA+U calculations for the experimental layered structure of \CFMO. The measured lattice constants were used,\cite{Hosaka15} and the atomic positions were optimized. To model the observed Fe/Mn layered order and the CD (2Fe$^{4+}$ $\rightarrow$ Fe$^{3+}$ + Fe$^{5+}$), we initialized the corresponding density matrix $t_{2g}^3e_g^2$, $t_{2g}^3$, and $t_{2g}^3$ for Fe$^{3+}$, Fe$^{5+}$, and Mn$^{4+}$ ions, respectively. All the nearest neighboring magnetic couplings are treated antiferromagnetic.
With full electronic and atomic relaxations, we indeed achieved the stable CD solution with layered Fe/Mn ordering and with CD in the Fe layers.
This CD solution has two different sets of Fe-O bondlengths, one with the in-plane 2.009 \AA $\times$ 4 and out-of-plane 1.957 \AA $\times$ 2, and the other with 1.840 \AA $\times$ 4 and 1.925 \AA $\times$ 2. (Note that as the CD occurs in the Fe$^{4+}$ layers of {\CFMO}, the in-plane oxygens displace a lot to accommodate the much different planar Fe$^{3+}$-O and Fe$^{5+}$-O bonds. However, the out-of-plane oxygens shift much less as their strong covalency with the neighboring Mn$^{4+}$ ions, via the $pd\sigma$ hybridization of the empty Mn$^{4+}$ $e_g$ orbital, do not allow themselves to move a lot.) The set of longer Fe-O bonds corresponds to Fe$^{3+}$ and the other set of shorter ones to Fe$^{5+}$. This assignment is supported by the larger ionic size of Fe$^{3+}$ (0.645 \AA) $vs$ the smaller one of Fe$^{5+}$ (0.525 \AA),\cite{Shannon} and their size difference of 0.12 \AA~ perfectly matches the average bondlength difference between Fe$^{3+}$-O (1.992 \AA) and Fe$^{5+}$-O (1.868 \AA). In addition, these two average bondlengths are very close to those in the CD CaFeO$_3$ with 1.974 \AA~ for Fe$^{3+}$-O and 1.872 \AA~ for Fe$^{5+}$-O.\cite{Woodward}  In contrast to the CD Fe ions, the Mn$^{4+}$ ions in the layer ordered {\CFMO} have very similar Mn-O average bondlengths: 1.910 \AA$\times$6 (in-plane 1.931 \AA$\times$4 and out-of-plane 1.869 \AA$\times$2) for the Mn$^{4+}$ neighboring to the Fe$^{3+}$, and 1.920 \AA$\times$6
(1.913 \AA$\times$4 and 1.935 \AA$\times$2) for the Mn$^{4+}$ neighboring to the Fe$^{5+}$, see Fig. 1. Moreover, the Fe$^{3+}$ ion has a local spin moment of 4.01 $\mu_{\rm B}$ within the muffin tin sphere, and Fe$^{5+}$ 2.46 $\mu_{\rm B}$, see Table I. A certain reduction from their respective formal spin Fe$^{3+}$ $S$=5/2 and Fe$^{5+}$ $S$=3/2 is due to the strong covalency with the ligand oxygens. Correspondingly, the formal $S$=3/2 Mn$^{4+}$ ion has a reduced spin moment of 2.61 $\mu_{\rm B}$.

 \begin{figure}[t]
\includegraphics[width=9.5cm]{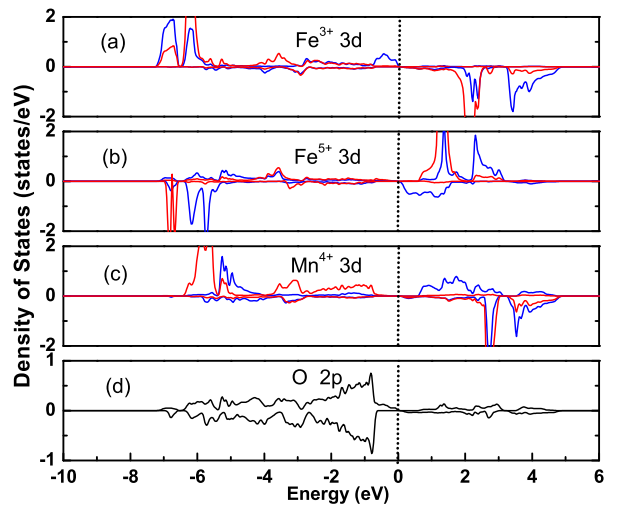}
 \caption{The $3d$-orbital DOS of (a) Fe$^{3+}$, (b) Fe$^{5+}$, and (c) Mn$^{4+}$, and (d) O $2p$ DOS of the layered {\CFMO} by LSDA+U. The blue (red) curves stand for the $e_{g}$ ($t_{2g}$) state. The positive (negative) value stands for the up (down) spin channel. Fermi level is set at zero energy.
}
%\label{fig2:el}
\end{figure}
\begin{table}[t]
  \caption{Relative total energies $\Delta$E (meV/fu) and local spin moments ($\mu$$_{\rm B}$) of the CD layered, the rock salt, and the CD rock salt structures of {\CFMO} calculated by LSDA+U. The CD rock salt structure assumes a Fe$^{3+}$-Fe$^{5+}$ CD with two different average Fe-O bondlengths adapted from the CD layered structure.
  The corresponding data for the fully relaxed structures (both the volume and the atomic positions) are listed in the round brackets. The data for the hypothetical Ca$_{2}$TiMnO$_{6}$ are also included.
}
  \label{tb1}
  \begin{tabular}{l@{\hskip5mm}c@{\hskip5mm}c@{\hskip5mm}c@{\hskip5mm}c@{\hskip5mm}}%c@{\hskip5mm}c}%c@{\hskip5mm}c}
\hline\hline
   Ca$_{2}$FeMnO$_{6}$ & $\Delta$E & Fe$^{3+}$ & Fe$^{5+}$ & Mn$^{4+}$ \\ \hline
    CD layered              & 0     & 4.01 & 2.46 & 2.61  \\
                       & (0)     & (3.97) & (2.33) & (2.57)   \\ \hline
    rock salt         & 78   & \multicolumn{2}{c}{3.60 Fe$^{4+}$} & 2.48  \\
                       & (99) & \multicolumn{2}{c}{(3.50 Fe$^{4+}$)} & (2.44)\\ \hline
    CD rock salt         & 221 & 3.89 & 3.03 & 2.50  \\
                       & (260) & (3.85) & (2.91) & (2.46)\\
\hline\hline
   Ca$_{2}$TiMnO$_{6}$ & $\Delta$E & Ti$^{4+}$  & Mn$^{4+}$ \\ \hline
    layered              & 0     & 0.00 &  2.68  \\
                       & (0)     & (0.00)  & (2.63)   \\ \hline
    rock salt         & --113   & 0.00 &  2.68  \\
                       & (--117) & (0.00) &  (2.64)\\
\hline\hline
 \end{tabular}
\end{table}

We plot in Fig. 2 the orbitally resolved density of states (DOS) of {\CFMO}. The LSDA+U calculations give an insulating solution for the layered CD state. It can be seen in Fig. 2(a) that the formal Fe$^{3+}$ ion has, as expected for its $t_{2g}^3e_g^2$ $S$=5/2 configuration, the fully occupied majority-spin $t_{2g}$ and $e_g$ orbitals, but the minority-spin ones are fully unoccupied. Due to the strong covalency with the ligand oxygens, the lower-energy bonding state at 7 eV below the Fermi level has an even more $e_g$ component than $t_{2g}$, although $e_g$ is a higher crystal field level than $t_{2g}$. In contrast, the formal Fe$^{5+}$ ion has the occupied majority-spin $t_{2g}$ orbital as seen in Fig. 2(b), but the majority-spin $e_g$ is only partially occupied and this partial occupation is due to the bonding state with the ligand oxygens. Therefore, one could say the Fe$^{5+}$ $e_g$ state is formally unoccupied but it gains some occupation due to a very strong covalency with the ligand oxygens as supported by the shortest Fe$^{5+}$-O bonds. Moreover, the negative charge transfer character of the unusual high valent Fe$^{5+}$ ion favors this strong covalency to form the actual Fe$^{3+}\underline{L}^2$ state rather than the nominal Fe$^{5+}$ (both the states have the same formal $S$=3/2). By a comparison between Figs. 2(a) and 2(b), one can see that the formal Fe$^{5+}$ has a lower $3d$ on-site energy (the center of gravity of the $3d$ DOS) than the Fe$^{3+}$, and this is more clear for the unoccupied minority-spin $3d$ states. This accords with the chemical trend that for a given transition metal, a higher valence state has a lower on-site energy than a lower valence due to an enhanced nuclear attraction in the former. Fig. 2(c) shows that the Mn$^{4+}$ ion has a fully occupied majority-spin $t_{2g}$ orbital but a partial $e_g$ occupation due to a strong covalency of the formally unoccupied $e_g$ orbital with the ligand oxygens as above for the Fe$^{5+}$. Therefore, the Mn$^{4+}$ ion is in the formal $t_{2g}^3$ $S$=3/2 state. Furthermore, one can see in Fig. 2(d) that the O $2p$ state has a largest contribution in the topmost valence bands, which reflects the charge transfer character of the insulating gap in this negative charge transfer oxide with an unusual high valence.\cite{Khomskii,Zaanen,Bocquet}

Now we turn to the possible rock salt structure of {\CFMO} with a checkerboard arrangement of the Fe and Mn ions, which is a common structure in the B-site ordered double perovskite. In order to make a direct comparison with the above CD layered order, here we use a cubic $2\times2\times2$ supercell for the rock salt structure with a same lattice volume as above, and the interior atomic positions are relaxed. It is interesting that independent of the initialized Fe$^{4+}$-Mn$^{4+}$ state or the Fe CD one as done in the LSDA+U calculations for this rock salt structure, both the states converge to an exactly same Fe$^{4+}$-Mn$^{4+}$ state after a full electronic and atomic relaxation. The Fe$^{4+}$-Mn$^{4+}$ state in the rock salt structure has a uniform Fe-O bondlength of 1.940 \AA$\times$6, just in between the calculated  1.992 \AA~ for Fe$^{3+}$-O and 1.868 \AA~ for Fe$^{5+}$-O in the above CD layered structure. The Mn$^{4+}$-O bondlength remains almost unchanged, 1.902 \AA$\times$6 here vs 1.910-1.920 \AA$\times$6 in the above CD layered structure. Moreover, the Fe$^{4+}$-Mn$^{4+}$ state has a local spin moment of 3.60 $\mu_{\rm B}$/Fe and 2.48 $\mu_{\rm B}$/Mn, which are reduced by a covalency from the formal $S$=2 for Fe$^{4+}$ ($t_{2g}^3e_g^1$) and $S$=3/2 for Mn$^{4+}$ ($t_{2g}^3$).

 \begin{figure}[t]
\includegraphics[width=9cm]{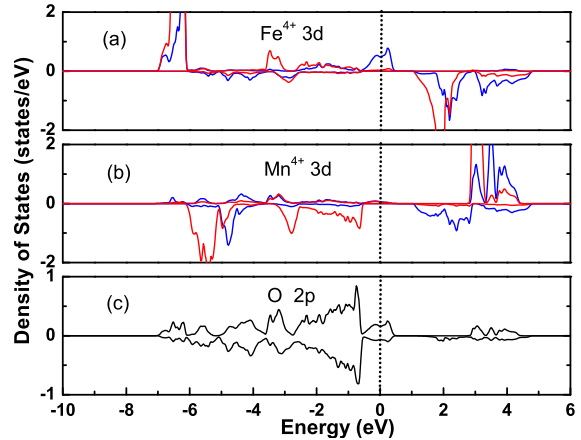}
 \caption{The $3d$-orbital DOS of (a) Fe$^{4+}$ and (b) Mn$^{4+}$, and (c) O $2p$ DOS of the rock salt structure {\CFMO} by LSDA+U. The blue (red) curves stand for the $e_{g}$ ($t_{2g}$) state. The positive (negative) value stands for the up (down) spin channel. Fermi level is set at zero energy.
}
%\label{fig2:el}
\end{figure}

The Fe$^{4+}$-Mn$^{4+}$ state in the rock salt structure is metallic due to the $e_g$ half filling of the Fe$^{4+}$ ($t_{2g}^3e_g^1$) ions, see the DOS results in Fig. 3. This accounts for the uniform Fe$^{4+}$ state with an electron itineracy but not a localized Fe$^{3+}$-Fe$^{5+}$ CD state. These calculations indicate that it is difficult and practically impossible to stabilize the CD phase for the rock salt Fe-Mn ordering (see more discussion below). Indeed, the total energy results show that the Fe$^{4+}$-Mn$^{4+}$ state in the rock salt structure lies higher in energy than the above CD layered structure by 78 meV/fu as seen in Table I.

 \begin{figure}[t]
\includegraphics[width=6cm]{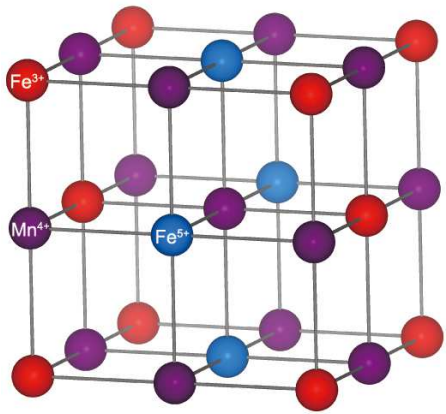}
 \caption{A schemetic illustration of the hypothetical rock salt structure {\CFMO} with a possible Fe$^{3+}$-Fe$^{5+}$ CD in the fcc sublattice. Such a CD would suffer a geometry frustration.
 }
%\label{fig2:el}
\end{figure}

To further confirm that the CD state in the hypothetic rock salt ordered {\CFMO} is unstable, we constructed a cubic $2\times2\times2$ supercell with the fixed (here no atomic relaxation) Fe$^{3+}$-O bondlength of 1.992 \AA$\times$6 and the Fe$^{5+}$-O one of 1.868 \AA$\times$6 (as adapted from the above CD layered structure). This mimics a Fe CD in the Fe-Mn checkerboard structure, and then the Fe$^{3+}$-Fe$^{5+}$ CD ions appear in each Fe-Mn layer, see an illustrative picture in Fig. 4. As seen in Table I, our LSDA+U calculation can stabilize this state (named CD rock salt in Table I), which has the spin moment of 3.89 $\mu_{\rm B}$ for Fe$^{3+}$, 3.03 $\mu_{\rm B}$ for Fe$^{5+}$, and 2.50 $\mu_{\rm B}$ for Mn$^{4+}$. Although this result signals a possible Fe CD in the Fe-Mn checkerboard structure, the corresponding total energy is much higher than the above CD layered structure by 221 meV/fu, once again showing the instability of the rock salt structure with the Fe CD. Note that when the atomic relaxation was carried out, this Fe CD in the rock salt structure disappears and evolves into the common rock salt structure with the uniform Fe$^{4+}$-O bonds, which lies higher in energy than the CD layered structure by 78 meV/fu as reported above (see Table I).

In order to substantiate the above conclusion and to confirm the close relationship between the unusual layered order of \CFMO~ and the Fe$^{4+}$ CD, here we carry out a comparative study for the isostructural \CTMO~ but without Fe$^{4+}$ ions leading to CD. We repeated the above calculations as listed in Table I but replaced Fe by Ti free of CD, both of which have a similar ionic size (Ti$^{4+}$ 0.605 \AA~ vs Fe$^{4+}$ 0.585 \AA~).\cite{Shannon} As seen in Table I, \CTMO~ has the invariant Ti$^{4+}$ and Mn$^{4+}$ state in both the layered structure and the rock salt one. The Ti$^{4+}$ ion is nonmagnetic and its calculated spin moment is 0, and the Mn$^{4+}$ ion has a local spin moment of 2.68 $\mu_{\rm B}$ representing its formal $S$=3/2. It is clear that the Ti$^{4+}$ ions have no CD solution. It is important to note that here the standard rock salt structure of \CTMO~ is more stable than the layered structure by 113 meV/fu, as seen in Table I. This result is completely opposite to the above one for \CFMO, which has the unusual layered order with the Fe CD. Therefore, this comparative study gives another strong indication that the unusual layered ordering of \CFMO~ is intrinsically connected with the occurrence of the Fe$^{3+}$-Fe$^{5+}$ CD of the formal Fe$^{4+}$ ions in {\CFMO}.

For completeness, and to be sure that we indeed obtained the real ground state of {\CFMO}, we carried out another LSDA+U calculation with a full structural optimization in the layered CD structure. The optimized lattice constants $a$=7.350 \AA, $b$=7.390 \AA~ and $c$=7.390 \AA~ (in the 2$\times$2$\times$2 supercell of the ABO$_{3}$ perovskite type, see Fig. 1) agree well with the experimental ones ($a$=7.495 \AA, $b$=7.489 \AA~ and $c$=7.519 \AA~)\cite{Hosaka15} within the typical error bar of $\pm$3\% given by density functional calculations. Corresponding calculations were also performed for the rock salt structure and the CD rock salt structure, and the obtained results are summarized in the round brackets in Table I. It is clear that our conclusion is robust and there are only insignificant numerical changes due to the full lattice and atomic relaxations. For example, the local spin moments of the formal Fe$^{3+}$ and Fe$^{5+}$ in the CD layered structure are now 3.97 $\mu_{\rm B}$ and 2.33 $\mu_{\rm B}$, respectively (see Table I for a comparison). Moreover, our calculations again find the rock salt structure to be less stable than the CD layered one of {\CFMO} by 99 meV/fu, and the opposite is true for Ca$_{2}$TiMnO$_{6}$. Therefore the above discussion and conclusion remain unchanged: evidently the unusual layered ordering of \CFMO~ goes hand in hand with the occurrence of the Fe$^{3+}$-Fe$^{5+}$ CD of the formal Fe$^{4+}$ ions which,  however would be frustrated in the conventional rock salt structure. Apparently this is responsible for the appearance of this unusual layered ordering in this double perovskite.

%$\times$
%%Summary
\section{\label{sec:sum}Summary}

All the above calculations prove that the unusual layered order of \CFMO~ with the Fe CD is the ground state. {\it Why does \CFMO~ have the unusual layer order rather than a common rock salt structure?}
This is because \CFMO~ has the nominal Fe$^{4+}$-Mn$^{4+}$ state. Although Mn$^{4+}$ is stable and has a closed $t_{2g}^3$ subshell, Fe$^{4+}$ ($t_{2g}^3e_g^1$) is an intrinsically unstable high valent Jahn-Teller ion with the formally half filled $e_g$ orbital. The Fe$^{4+}$ ions tend to undergo a CD into Fe$^{3+}$ ($t_{2g}^3e_g^2$) and Fe$^{5+}$ ($t_{2g}^3$) both have a formally closed subshell. In the standard rock salt structure, each Fe$^{4+}$ is surrounded by six Mn$^{4+}$ and vice versa. As a result, the Fe$^{4+}$ ions form a fcc sublattice.
If a CD occurs (see Fig. 4), the consequent Fe$^{3+}$ and Fe$^{5+}$ ions would suffer a serious frustration as in the well known antiferromagnetically coupled fcc lattice (here the two different Fe$^{3+}$ and Fe$^{5+}$ ions behave like the up and down spins in the antiferromagnetic fcc lattice). This should be the reason why the standard rock salt structure does not apply to the B-site ordered \CFMO. In contrast, in the unusual layered order of \CFMO, the nominal Fe$^{4+}$ and Mn$^{4+}$ ions form their respective layers (which alternate along the $a$ axis, see Fig. 1). Then each Fe$^{4+}$ layer can freely undergo a CD transition and form a planar Fe$^{3+}$-Fe$^{5+}$ checkerboard structure. This CD layered structure turns out to be the ground state as seen above. Therefore, all the above results give us more evidence that the unusual layered ordering of \CFMO~ is actually intrinsically connected with the tendency of the nominal Fe$^{4+}$ to CD into Fe$^{3+}$ and Fe$^{5+}$.

\section{\label{acknow}Acknowledgements}

K. Y. and H. W. were supported by the NSF of China (Grants No. 11474059 and No. 11674064) and by the National Key Research and Development Program of China (Grant No. 2016YFA0300700).
D. I. K. was supported by the Deutsche Forschungsgemeinschaft through CRC 1238.

\nocite{*}

\bibliography{Ca2FeMnO6_sub}% Produces the bibliography via BibTeX.

\begin{thebibliography}{19}%
\makeatletter
\providecommand \@ifxundefined [1]{%
 \@ifx{#1\undefined}
}%
\providecommand \@ifnum [1]{%
 \ifnum #1\expandafter \@firstoftwo
 \else \expandafter \@secondoftwo
 \fi
}%
\providecommand \@ifx [1]{%
 \ifx #1\expandafter \@firstoftwo
 \else \expandafter \@secondoftwo
 \fi
}%
\providecommand \natexlab [1]{#1}%
\providecommand \enquote  [1]{``#1''}%
\providecommand \bibnamefont  [1]{#1}%
\providecommand \bibfnamefont [1]{#1}%
\providecommand \citenamefont [1]{#1}%
\providecommand \href@noop [0]{\@secondoftwo}%
\providecommand \href [0]{\begingroup \@sanitize@url \@href}%
\providecommand \@href[1]{\@@startlink{#1}\@@href}%
\providecommand \@@href[1]{\endgroup#1\@@endlink}%
\providecommand \@sanitize@url [0]{\catcode `\\12\catcode `\$12\catcode
  `\&12\catcode `\#12\catcode `\^12\catcode `\_12\catcode `\%12\relax}%
\providecommand \@@startlink[1]{}%
\providecommand \@@endlink[0]{}%
\providecommand \url  [0]{\begingroup\@sanitize@url \@url }%
\providecommand \@url [1]{\endgroup\@href {#1}{\urlprefix }}%
\providecommand \urlprefix  [0]{URL }%
\providecommand \Eprint [0]{\href }%
\providecommand \doibase [0]{http://dx.doi.org/}%
\providecommand \selectlanguage [0]{\@gobble}%
\providecommand \bibinfo  [0]{\@secondoftwo}%
\providecommand \bibfield  [0]{\@secondoftwo}%
\providecommand \translation [1]{[#1]}%
\providecommand \BibitemOpen [0]{}%
\providecommand \bibitemStop [0]{}%
\providecommand \bibitemNoStop [0]{.\EOS\space}%
\providecommand \EOS [0]{\spacefactor3000\relax}%
\providecommand \BibitemShut  [1]{\csname bibitem#1\endcsname}%
\let\auto@bib@innerbib\@empty
%</preamble>
\bibitem [{\citenamefont {Tokura}\ and\ \citenamefont
  {Nagaosa}(2000)}]{Tokura}%
  \BibitemOpen
  \bibfield  {author} {\bibinfo {author} {\bibfnamefont {Y.}~\bibnamefont
  {Tokura}}\ and\ \bibinfo {author} {\bibfnamefont {N.}~\bibnamefont
  {Nagaosa}},\ }\href@noop {} {\bibfield  {journal} {\bibinfo  {journal}
  {Science}\ }\textbf {\bibinfo {volume} {288}},\ \bibinfo {pages} {462}
  (\bibinfo {year} {2000})}\BibitemShut {NoStop}%
\bibitem [{\citenamefont {Khomskii}(2014)}]{Khomskii}%
  \BibitemOpen
  \bibfield  {author} {\bibinfo {author} {\bibfnamefont {D.~I.}\ \bibnamefont
  {Khomskii}},\ }\href@noop {} {\emph {\bibinfo {title} {Transition Metal
  Compounds}}}\ (\bibinfo  {publisher} {Cambridge University Press},\ \bibinfo
  {year} {2014})\BibitemShut {NoStop}%
\bibitem [{\citenamefont {Alonso}\ \emph {et~al.}(1999)\citenamefont {Alonso},
  \citenamefont {Garc\'{\i}a-Mu\~noz}, \citenamefont {Fern\'andez-D\'{\i}az},
  \citenamefont {Aranda}, \citenamefont {Mart\'{\i}nez-Lope},\ and\
  \citenamefont {Casais}}]{Alonso}%
  \BibitemOpen
  \bibfield  {author} {\bibinfo {author} {\bibfnamefont {J.~A.}\ \bibnamefont
  {Alonso}}, \bibinfo {author} {\bibfnamefont {J.~L.}\ \bibnamefont
  {Garc\'{\i}a-Mu\~noz}}, \bibinfo {author} {\bibfnamefont {M.~T.}\
  \bibnamefont {Fern\'andez-D\'{\i}az}}, \bibinfo {author} {\bibfnamefont
  {M.~A.~G.}\ \bibnamefont {Aranda}}, \bibinfo {author} {\bibfnamefont {M.~J.}\
  \bibnamefont {Mart\'{\i}nez-Lope}}, \ and\ \bibinfo {author} {\bibfnamefont
  {M.~T.}\ \bibnamefont {Casais}},\ }\href@noop {} {\bibfield  {journal}
  {\bibinfo  {journal} {Phys. Rev. Lett.}\ }\textbf {\bibinfo {volume} {82}},\
  \bibinfo {pages} {3871} (\bibinfo {year} {1999})}\BibitemShut {NoStop}%
\bibitem [{\citenamefont {Mazin}\ \emph {et~al.}(2007)\citenamefont {Mazin},
  \citenamefont {Khomskii}, \citenamefont {Lengsdorf}, \citenamefont {Alonso},
  \citenamefont {Marshall}, \citenamefont {Ibberson}, \citenamefont
  {Podlesnyak}, \citenamefont {Mart\'{\i}nez-Lope},\ and\ \citenamefont
  {Abd-Elmeguid}}]{Mazin}%
  \BibitemOpen
  \bibfield  {author} {\bibinfo {author} {\bibfnamefont {I.~I.}\ \bibnamefont
  {Mazin}}, \bibinfo {author} {\bibfnamefont {D.~I.}\ \bibnamefont {Khomskii}},
  \bibinfo {author} {\bibfnamefont {R.}~\bibnamefont {Lengsdorf}}, \bibinfo
  {author} {\bibfnamefont {J.~A.}\ \bibnamefont {Alonso}}, \bibinfo {author}
  {\bibfnamefont {W.~G.}\ \bibnamefont {Marshall}}, \bibinfo {author}
  {\bibfnamefont {R.~M.}\ \bibnamefont {Ibberson}}, \bibinfo {author}
  {\bibfnamefont {A.}~\bibnamefont {Podlesnyak}}, \bibinfo {author}
  {\bibfnamefont {M.~J.}\ \bibnamefont {Mart\'{\i}nez-Lope}}, \ and\ \bibinfo
  {author} {\bibfnamefont {M.~M.}\ \bibnamefont {Abd-Elmeguid}},\ }\href@noop
  {} {\bibfield  {journal} {\bibinfo  {journal} {Phys. Rev. Lett.}\ }\textbf
  {\bibinfo {volume} {98}},\ \bibinfo {pages} {176406} (\bibinfo {year}
  {2007})}\BibitemShut {NoStop}%
\bibitem [{\citenamefont {Kawasaki}\ \emph {et~al.}(1998)\citenamefont
  {Kawasaki}, \citenamefont {Takano}, \citenamefont {Kanno}, \citenamefont
  {Takeda},\ and\ \citenamefont {Fujimori}}]{Kawasaki}%
  \BibitemOpen
  \bibfield  {author} {\bibinfo {author} {\bibfnamefont {S.}~\bibnamefont
  {Kawasaki}}, \bibinfo {author} {\bibfnamefont {M.}~\bibnamefont {Takano}},
  \bibinfo {author} {\bibfnamefont {R.}~\bibnamefont {Kanno}}, \bibinfo
  {author} {\bibfnamefont {T.}~\bibnamefont {Takeda}}, \ and\ \bibinfo {author}
  {\bibfnamefont {A.}~\bibnamefont {Fujimori}},\ }\href@noop {} {\bibfield
  {journal} {\bibinfo  {journal} {J. Phys. Soc. Jpn.}\ }\textbf {\bibinfo
  {volume} {67}},\ \bibinfo {pages} {1529} (\bibinfo {year}
  {1998})}\BibitemShut {NoStop}%
\bibitem [{\citenamefont {Woodward}\ \emph {et~al.}(2000)\citenamefont
  {Woodward}, \citenamefont {Cox}, \citenamefont {Moshopoulou}, \citenamefont
  {Sleight},\ and\ \citenamefont {Morimoto}}]{Woodward}%
  \BibitemOpen
  \bibfield  {author} {\bibinfo {author} {\bibfnamefont {P.~M.}\ \bibnamefont
  {Woodward}}, \bibinfo {author} {\bibfnamefont {D.~E.}\ \bibnamefont {Cox}},
  \bibinfo {author} {\bibfnamefont {E.}~\bibnamefont {Moshopoulou}}, \bibinfo
  {author} {\bibfnamefont {A.~W.}\ \bibnamefont {Sleight}}, \ and\ \bibinfo
  {author} {\bibfnamefont {S.}~\bibnamefont {Morimoto}},\ }\href@noop {}
  {\bibfield  {journal} {\bibinfo  {journal} {Phys. Rev. B}\ }\textbf {\bibinfo
  {volume} {62}},\ \bibinfo {pages} {844} (\bibinfo {year} {2000})}\BibitemShut
  {NoStop}%
\bibitem [{\citenamefont {Yamada}\ \emph {et~al.}(2008)\citenamefont {Yamada},
  \citenamefont {Takata}, \citenamefont {Hayashi}, \citenamefont {Shinohara},
  \citenamefont {Azuma}, \citenamefont {Mori}, \citenamefont {Muranaka},
  \citenamefont {Shimakawa},\ and\ \citenamefont {Takano}}]{Yamada}%
  \BibitemOpen
  \bibfield  {author} {\bibinfo {author} {\bibfnamefont {I.}~\bibnamefont
  {Yamada}}, \bibinfo {author} {\bibfnamefont {K.}~\bibnamefont {Takata}},
  \bibinfo {author} {\bibfnamefont {N.}~\bibnamefont {Hayashi}}, \bibinfo
  {author} {\bibfnamefont {S.}~\bibnamefont {Shinohara}}, \bibinfo {author}
  {\bibfnamefont {M.}~\bibnamefont {Azuma}}, \bibinfo {author} {\bibfnamefont
  {S.}~\bibnamefont {Mori}}, \bibinfo {author} {\bibfnamefont {S.}~\bibnamefont
  {Muranaka}}, \bibinfo {author} {\bibfnamefont {Y.}~\bibnamefont {Shimakawa}},
  \ and\ \bibinfo {author} {\bibfnamefont {M.}~\bibnamefont {Takano}},\
  }\href@noop {} {\bibfield  {journal} {\bibinfo  {journal} {Angew. Chem. Int.
  Ed.}\ }\textbf {\bibinfo {volume} {47}},\ \bibinfo {pages} {7032} (\bibinfo
  {year} {2008})}\BibitemShut {NoStop}%
\bibitem [{\citenamefont {Mizumaki}\ \emph {et~al.}(2011)\citenamefont
  {Mizumaki}, \citenamefont {Chen}, \citenamefont {Saito}, \citenamefont
  {Yamada}, \citenamefont {Attfield},\ and\ \citenamefont
  {Shimakawa}}]{Mizumaki}%
  \BibitemOpen
  \bibfield  {author} {\bibinfo {author} {\bibfnamefont {M.}~\bibnamefont
  {Mizumaki}}, \bibinfo {author} {\bibfnamefont {W.~T.}\ \bibnamefont {Chen}},
  \bibinfo {author} {\bibfnamefont {T.}~\bibnamefont {Saito}}, \bibinfo
  {author} {\bibfnamefont {I.}~\bibnamefont {Yamada}}, \bibinfo {author}
  {\bibfnamefont {J.~P.}\ \bibnamefont {Attfield}}, \ and\ \bibinfo {author}
  {\bibfnamefont {Y.}~\bibnamefont {Shimakawa}},\ }\href@noop {} {\bibfield
  {journal} {\bibinfo  {journal} {Phys. Rev. B}\ }\textbf {\bibinfo {volume}
  {84}},\ \bibinfo {pages} {094418} (\bibinfo {year} {2011})}\BibitemShut
  {NoStop}%
\bibitem [{\citenamefont {Zaanen}\ \emph {et~al.}(1985)\citenamefont {Zaanen},
  \citenamefont {Sawatzky},\ and\ \citenamefont {Allen}}]{Zaanen}%
  \BibitemOpen
  \bibfield  {author} {\bibinfo {author} {\bibfnamefont {J.}~\bibnamefont
  {Zaanen}}, \bibinfo {author} {\bibfnamefont {G.~A.}\ \bibnamefont
  {Sawatzky}}, \ and\ \bibinfo {author} {\bibfnamefont {J.~W.}\ \bibnamefont
  {Allen}},\ }\href@noop {} {\bibfield  {journal} {\bibinfo  {journal} {Phys.
  Rev. Lett.}\ }\textbf {\bibinfo {volume} {55}},\ \bibinfo {pages} {418}
  (\bibinfo {year} {1985})}\BibitemShut {NoStop}%
\bibitem [{\citenamefont {Bocquet}\ \emph {et~al.}(1992)\citenamefont
  {Bocquet}, \citenamefont {Fujimori}, \citenamefont {Mizokawa}, \citenamefont
  {Saitoh}, \citenamefont {Namatame}, \citenamefont {Suga}, \citenamefont
  {Kimizuka}, \citenamefont {Takeda},\ and\ \citenamefont {Takano}}]{Bocquet}%
  \BibitemOpen
  \bibfield  {author} {\bibinfo {author} {\bibfnamefont {A.~E.}\ \bibnamefont
  {Bocquet}}, \bibinfo {author} {\bibfnamefont {A.}~\bibnamefont {Fujimori}},
  \bibinfo {author} {\bibfnamefont {T.}~\bibnamefont {Mizokawa}}, \bibinfo
  {author} {\bibfnamefont {T.}~\bibnamefont {Saitoh}}, \bibinfo {author}
  {\bibfnamefont {H.}~\bibnamefont {Namatame}}, \bibinfo {author}
  {\bibfnamefont {S.}~\bibnamefont {Suga}}, \bibinfo {author} {\bibfnamefont
  {N.}~\bibnamefont {Kimizuka}}, \bibinfo {author} {\bibfnamefont
  {Y.}~\bibnamefont {Takeda}}, \ and\ \bibinfo {author} {\bibfnamefont
  {M.}~\bibnamefont {Takano}},\ }\href@noop {} {\bibfield  {journal} {\bibinfo
  {journal} {Phys. Rev. B}\ }\textbf {\bibinfo {volume} {45}},\ \bibinfo
  {pages} {1561} (\bibinfo {year} {1992})}\BibitemShut {NoStop}%
\bibitem [{\citenamefont {Hosaka}\ \emph {et~al.}(2017)\citenamefont {Hosaka},
  \citenamefont {Romero}, \citenamefont {Ichikawa}, \citenamefont {Saito},\
  and\ \citenamefont {Shimakawa}}]{Hosaka17}%
  \BibitemOpen
  \bibfield  {author} {\bibinfo {author} {\bibfnamefont {Y.}~\bibnamefont
  {Hosaka}}, \bibinfo {author} {\bibfnamefont {F.~D.}\ \bibnamefont {Romero}},
  \bibinfo {author} {\bibfnamefont {N.}~\bibnamefont {Ichikawa}}, \bibinfo
  {author} {\bibfnamefont {T.}~\bibnamefont {Saito}}, \ and\ \bibinfo {author}
  {\bibfnamefont {Y.}~\bibnamefont {Shimakawa}},\ }\href@noop {} {\bibfield
  {journal} {\bibinfo  {journal} {Angew. Chem. Int. Ed.}\ }\textbf {\bibinfo
  {volume} {56}},\ \bibinfo {pages} {4243} (\bibinfo {year}
  {2017})}\BibitemShut {NoStop}%
\bibitem [{\citenamefont {Hosaka}\ \emph {et~al.}(2015)\citenamefont {Hosaka},
  \citenamefont {Ichikawa}, \citenamefont {Saito}, \citenamefont {Manuel},
  \citenamefont {Khalyavin}, \citenamefont {Attfield},\ and\ \citenamefont
  {Shimakawa}}]{Hosaka15}%
  \BibitemOpen
  \bibfield  {author} {\bibinfo {author} {\bibfnamefont {Y.}~\bibnamefont
  {Hosaka}}, \bibinfo {author} {\bibfnamefont {N.}~\bibnamefont {Ichikawa}},
  \bibinfo {author} {\bibfnamefont {T.}~\bibnamefont {Saito}}, \bibinfo
  {author} {\bibfnamefont {P.}~\bibnamefont {Manuel}}, \bibinfo {author}
  {\bibfnamefont {D.}~\bibnamefont {Khalyavin}}, \bibinfo {author}
  {\bibfnamefont {J.~P.}\ \bibnamefont {Attfield}}, \ and\ \bibinfo {author}
  {\bibfnamefont {Y.}~\bibnamefont {Shimakawa}},\ }\href@noop {} {\bibfield
  {journal} {\bibinfo  {journal} {J. Am. Chem. Soc.}\ }\textbf {\bibinfo
  {volume} {137}},\ \bibinfo {pages} {7468} (\bibinfo {year}
  {2015})}\BibitemShut {NoStop}%
\bibitem [{\citenamefont {Mizokawa}\ \emph {et~al.}(2000)\citenamefont
  {Mizokawa}, \citenamefont {Khomskii},\ and\ \citenamefont
  {Sawatzky}}]{Mizokawa}%
  \BibitemOpen
  \bibfield  {author} {\bibinfo {author} {\bibfnamefont {T.}~\bibnamefont
  {Mizokawa}}, \bibinfo {author} {\bibfnamefont {D.~I.}\ \bibnamefont
  {Khomskii}}, \ and\ \bibinfo {author} {\bibfnamefont {G.~A.}\ \bibnamefont
  {Sawatzky}},\ }\href@noop {} {\bibfield  {journal} {\bibinfo  {journal}
  {Phys. Rev. B}\ }\textbf {\bibinfo {volume} {61}},\ \bibinfo {pages} {11263}
  (\bibinfo {year} {2000})}\BibitemShut {NoStop}%
\bibitem [{\citenamefont {Johnston}\ \emph {et~al.}(2014)\citenamefont
  {Johnston}, \citenamefont {Mukherjee}, \citenamefont {Elfimov}, \citenamefont
  {Berciu},\ and\ \citenamefont {Sawatzky}}]{Johnston}%
  \BibitemOpen
  \bibfield  {author} {\bibinfo {author} {\bibfnamefont {S.}~\bibnamefont
  {Johnston}}, \bibinfo {author} {\bibfnamefont {A.}~\bibnamefont {Mukherjee}},
  \bibinfo {author} {\bibfnamefont {I.}~\bibnamefont {Elfimov}}, \bibinfo
  {author} {\bibfnamefont {M.}~\bibnamefont {Berciu}}, \ and\ \bibinfo {author}
  {\bibfnamefont {G.~A.}\ \bibnamefont {Sawatzky}},\ }\href@noop {} {\bibfield
  {journal} {\bibinfo  {journal} {Phys. Rev. Lett.}\ }\textbf {\bibinfo
  {volume} {112}},\ \bibinfo {pages} {106404} (\bibinfo {year}
  {2014})}\BibitemShut {NoStop}%
\bibitem [{\citenamefont {Anderson}\ \emph {et~al.}(1993)\citenamefont
  {Anderson}, \citenamefont {Greenwood}, \citenamefont {Taylor},\ and\
  \citenamefont {Poeppelmeier}}]{Anderson}%
  \BibitemOpen
  \bibfield  {author} {\bibinfo {author} {\bibfnamefont {M.}~\bibnamefont
  {Anderson}}, \bibinfo {author} {\bibfnamefont {K.}~\bibnamefont {Greenwood}},
  \bibinfo {author} {\bibfnamefont {G.}~\bibnamefont {Taylor}}, \ and\ \bibinfo
  {author} {\bibfnamefont {K.}~\bibnamefont {Poeppelmeier}},\ }\href@noop {}
  {\bibfield  {journal} {\bibinfo  {journal} {Prog. Solid State Chem.}\
  }\textbf {\bibinfo {volume} {22}},\ \bibinfo {pages} {197} (\bibinfo {year}
  {1993})}\BibitemShut {NoStop}%
\bibitem [{\citenamefont {Serrate}\ \emph {et~al.}(2007)\citenamefont
  {Serrate}, \citenamefont {Teresa},\ and\ \citenamefont {Ibarra}}]{Serrate}%
  \BibitemOpen
  \bibfield  {author} {\bibinfo {author} {\bibfnamefont {D.}~\bibnamefont
  {Serrate}}, \bibinfo {author} {\bibfnamefont {J.~M.~D.}\ \bibnamefont
  {Teresa}}, \ and\ \bibinfo {author} {\bibfnamefont {M.~R.}\ \bibnamefont
  {Ibarra}},\ }\href@noop {} {\bibfield  {journal} {\bibinfo  {journal} {J
  Phys-Condens Mat}\ }\textbf {\bibinfo {volume} {19}},\ \bibinfo {pages}
  {023201} (\bibinfo {year} {2007})}\BibitemShut {NoStop}%
\bibitem [{\citenamefont {Shannon}(1976)}]{Shannon}%
  \BibitemOpen
  \bibfield  {author} {\bibinfo {author} {\bibfnamefont {R.~D.}\ \bibnamefont
  {Shannon}},\ }\href@noop {} {\bibfield  {journal} {\bibinfo  {journal} {Acta
  Crystallogr A}\ }\textbf {\bibinfo {volume} {32}},\ \bibinfo {pages} {751}
  (\bibinfo {year} {1976})}\BibitemShut {NoStop}%
\bibitem [{\citenamefont {Blaha}\ \emph {et~al.}()\citenamefont {Blaha},
  \citenamefont {Schwarz}, \citenamefont {Madsen}, \citenamefont {Kvasnicka},\
  and\ \citenamefont {Luitz}}]{WIEN2K}%
  \BibitemOpen
  \bibfield  {author} {\bibinfo {author} {\bibfnamefont {P.}~\bibnamefont
  {Blaha}}, \bibinfo {author} {\bibfnamefont {K.}~\bibnamefont {Schwarz}},
  \bibinfo {author} {\bibfnamefont {G.}~\bibnamefont {Madsen}}, \bibinfo
  {author} {\bibfnamefont {D.}~\bibnamefont {Kvasnicka}}, \ and\ \bibinfo
  {author} {\bibfnamefont {J.}~\bibnamefont {Luitz}},\ }\href@noop {} {\enquote
  {\bibinfo {title} {Wien2k package},}\ }\bibinfo {howpublished}
  {http://www.wien2k.at}\BibitemShut {NoStop}%
\bibitem [{\citenamefont {Anisimov}\ \emph {et~al.}(1993)\citenamefont
  {Anisimov}, \citenamefont {Solovyev}, \citenamefont {Korotin}, \citenamefont
  {Czy\ifmmode~\dot{z}\else \.{z}\fi{}yk},\ and\ \citenamefont
  {Sawatzky}}]{Anisimov}%
  \BibitemOpen
  \bibfield  {author} {\bibinfo {author} {\bibfnamefont {V.~I.}\ \bibnamefont
  {Anisimov}}, \bibinfo {author} {\bibfnamefont {I.~V.}\ \bibnamefont
  {Solovyev}}, \bibinfo {author} {\bibfnamefont {M.~A.}\ \bibnamefont
  {Korotin}}, \bibinfo {author} {\bibfnamefont {M.~T.}\ \bibnamefont
  {Czy\ifmmode~\dot{z}\else \.{z}\fi{}yk}}, \ and\ \bibinfo {author}
  {\bibfnamefont {G.~A.}\ \bibnamefont {Sawatzky}},\ }\href@noop {} {\bibfield
  {journal} {\bibinfo  {journal} {Phys. Rev. B}\ }\textbf {\bibinfo {volume}
  {48}},\ \bibinfo {pages} {16929} (\bibinfo {year} {1993})}\BibitemShut
  {NoStop}%
\end{thebibliography}%
\end{document}